\documentclass[aps,pra,twocolumn,showpacs,superscriptaddress]{revtex4-1}
\usepackage{amsmath,dcolumn,bm, amsfonts, amssymb, graphicx,subfig}
\usepackage{caption}
\begin{document}
\title{Phonon induced two-mode squeezing of nitrogen-vacancy center ensembles}
\author{Qidong Xu}
\affiliation{Department of English, China Foreign Affair University, Beijing 100037, China}
\author{W. L. Yang} \email{ywl@wipm.ac.cn}
\affiliation{State Key Laboratory of Magnetic Resonance and Atomic and Molecular Physics, Wuhan Institute of Physics and Mathematics, Chinese Academy of Sciences, Wuhan 430071, China}
\author{Zhang-qi Yin} \email{yinzhangqi@mail.tsinghua.edu.cn}
\affiliation{Center for Quantum Information, Institute for Interdisciplinary Information Sciences, Tsinghua University, Beijing 100084, China}

\begin{abstract}
We propose a potentially practical scheme for realization of two-mode squeezed state with respect
to two distant nitrogen-vacancy center ensembles coupled to two interconnected mechanical modes of
diamond nanoresonators. By making use of the tunable phonon-spin interaction and the engineered
phonon-phonon tunneling, both the desired excitation transfer process and the optimal two-mode
squeezing between spin ensembles can be realized. We investigate the dynamics of the total system under infulences from the  mechanical decay using both analytical and
numerical methods, where the realistic conditions that leads to optimized squeezing between spin
ensembles are analyzed.

\end{abstract}
\date{\today}
\maketitle
\section{Introduction}
 Recent years have witnessed great advances in quantum information
technology. Efforts have been devoted to the quantum information
processing (QIP) based on quantum dots \cite{bennett2010strong}, atoms \cite%
{hammerer2009strong}, superconducting qubits \cite{devoret2013superconducting}, and nitrogen-vacancy (NV) center \cite{yin2015hybrid,Dujiangfeng,Zuchong,
andersen2012squeezing}. Among them, the atom-like NV centers in diamond have
emerged as a particular promising candidate for implementing quantum
technologies, since they exhibit long coherence time even at room
temperature. Besides, due to the collective excitation, an increasing of the
magnetic-dipole coupling and a strong coupling regime can be obtained
in the NV center ensemble (NVE) contained systems \cite{yang2012two,Lipengbo,song2015one,Yangwanli,Yangwanli1}. Benefited from the rapid
technical progress in the fabrication of diamond nanotructures \cite%
{ovartchaiyapong2012high,zalalutdinov2011ultrathin,burek2012free}, many
potential applications based on the mechanism of coupling a single NV
center \cite{2011,2015,Yin2013} or NVE to a mechanical mode have been studied both
theoretically and experimentally \cite{bennett2013phonon,macquarrie2013mechanical,albrecht2013coupling,goldman2015phonon,Optica}, such as cooling mechanical
resonator \cite{Xiaoyunfeng}, quantum interface \cite{Xuezhengyuan,Lipengbo1}, etc.

In this paper, we propose a new approach to generate a two-mode squeezed
state of two separate NVEs, which are embedded in two diamond nanoresonators.
The two diamond resonators are connected by a junction that gives rising to
a phonon-phonon interaction as shown in Fig.1. The basic idea of our work is
to first arrange all the NVs in one nanoresonator to be in the ground state
and the other resonator with all the NVs in the excited state. Those two
NVEs are separately coupled to the ground mechanical modes of the diamond
resonators, which are induced by the motion of them at low temperature \cite%
{maze2011properties,doherty2012theory}, {and the vibration due to the
ground mechanical mode of the nanoresonator changes the local strain where
the NV center is located, and results in an effective, strain-induced
electric field.} In this way we expect excitation transfer process from one
ensemble to the other via the phonon-spin interaction and the phonon-phonon
interaction. {In the present work we use the Holstein-Primakoff (HP)
approximation \cite{holstein1940field} to describe the spin ensembles in the
low-excitation regime,} and we find that the paired excitations could be
realized, which leads to the two-mode squeezed state of NVEs. {%
Meantime, the degree of squeezing can be manipulated and optimized easily}
in our model, by tuning the magnetic field or adjusting the phonon-phonon
interaction strength. This can actually save us the effort of precise time
control to detect the entanglement. We notice that our system can reach the
minimum of squeezing in a rather short time, so the mechanical dissipation
or the spin dephasing time will not cause considerable impact to our result.
{Our idea provides a scalable way to a NVE-based continuous-variable
QIP, which is close to being achievable with currently available technology.}

In Sec.\uppercase\expandafter{\romannumeral2} we describe the physical
systems in detail and give the mathematical model that we shall use through
out the paper. In Sec.\uppercase\expandafter{\romannumeral3}, using
adiabatic approximation to eliminate the mechanical mode, we compute the
expression of squeezing in the case of large detunings, and we confirm our
result by numerical simulation. In Sec.\uppercase\expandafter{\romannumeral4}
we investigate how to optimize the squeezing and give the {key}
parameters that lead to exponential decreasing squeezing. In Sec.\uppercase%
\expandafter{\romannumeral5} we discuss the realistic conditions and
conclude {the present work.}

\section{Model}

\begin{figure}[htbp]
\begin{centering}
\subfloat[]{
	\begin{centering}
	  \includegraphics[width =3in]{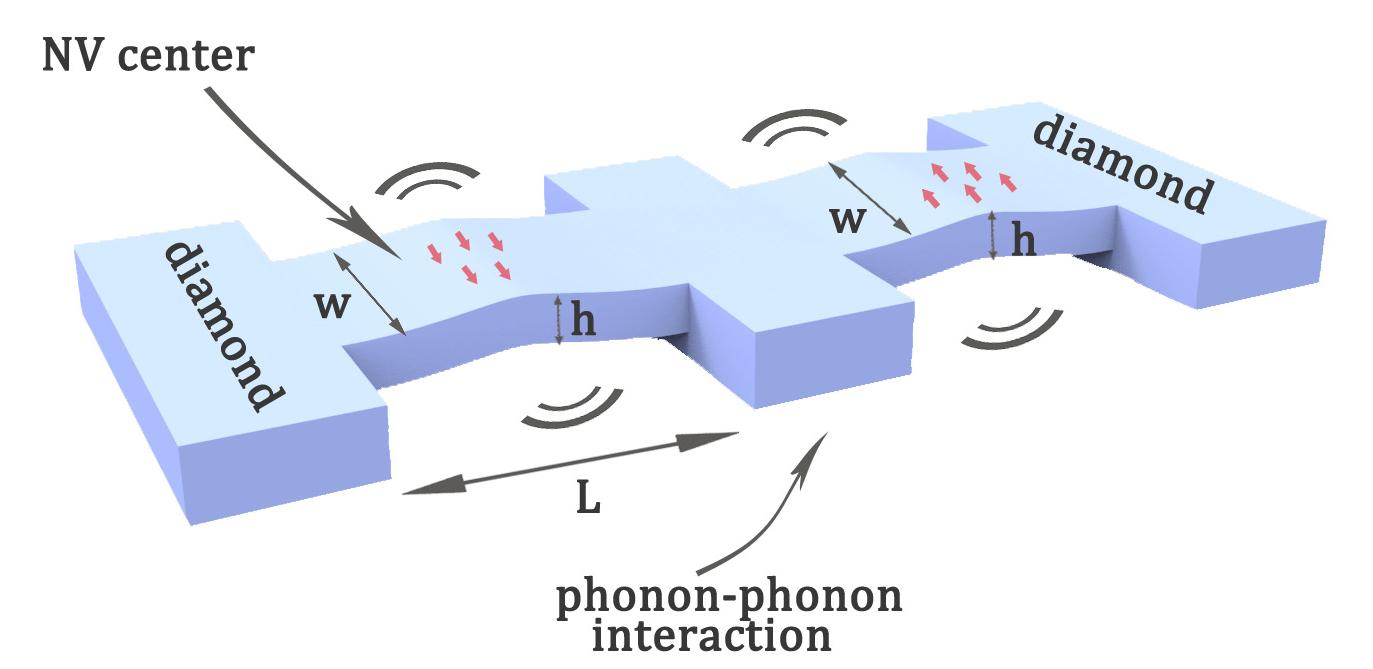}
	  \par\end{centering}

}

\subfloat[]{
    \begin{centering}
      \includegraphics[width=1.7in]{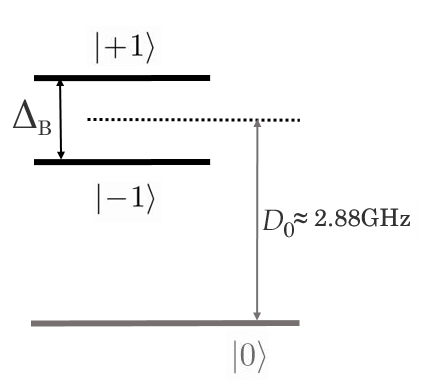}
      \par\end{centering}
}
\subfloat[]{
     \begin{centering}
       \includegraphics[width = 1.7in]{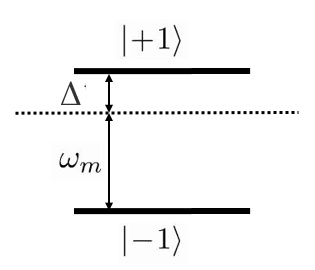}
       \par \end{centering}
}
\par\end{centering}
\captionsetup{justification = RaggedRight}
\label{fig1}

\caption{(a)Diamond clamped mechanical resonator with a junction in between, NVEs are separately embedded in the nanoresonators. (b) $^3A_2$ ground triplet states of the NV center with a single zero-field splitting of $D_0/2\pi\simeq 2.88 GHz$ between $m_s = 0$ and $m_s = \pm1$ spin sub-levels. (c) Spin in the two-level subspace ${|+1\rangle,|-1\rangle}$ is off resonatly coupled to a mechanical mode with detuning $\Delta$, which gives rising to the spin-phonon interaction. }
\end{figure}

We show the energy configuration of NV center in Fig.1 and each NV center is negatively charged. The degeneracy of the levels $m_s=\pm1$  which belongs to $^3A_2$ ground state can be breaked by an external magnetic field  with Zeeman splitting $\Delta_B=g_s\mu_BB_z/\hbar$, where $g_s\simeq2$ and $\mu_B$ is the Bohr magneton. On the other hand, the vibration due to the ground mechanical mode of the nanoresonator changes the local strain where the NV center is located, and gives rising to an effective, strain-induced electric field. The Hamiltonian for a single NV takes the form of($\hbar=1$)\cite{doherty2012theory}
\begin{align*}
H_{NV} = &(D_0 + d_{\|}E_z)S_z^2 + \mu_B g_s\vec{S}\cdot\vec{B} - \\
&d_{\bot}[E_x(S_xS_y+S_yS_x)+E_y(S_x^2-S_y^2)] \tag{1}
\end{align*}
where $D_0/2\pi\simeq 2.88 GHz$ is the zero field splitting, and $d_{\|}(d_{\bot})$ is the ground state electric dipole moment in the direction parallel(perpendicular) to the NV axis\cite{van1990electric, dolde2011electric}. As the strain  is linear to its position within small displacement, we have $E_{\bot}=E_0(a+a^\dagger )$ in which $a$ is the destruction operator of the mechanical mode and $E_0$ is the perpendicular strain resulted from the zero point motion of the beam. The $|\pm1\rangle$ states are both shifted from the ground state $|0\rangle$\cite{acosta2010temperature}. If we prepare all the NVs to be in the $|\pm1\rangle$ subspace at first and set the coupling of the mechanical mode of the nanobeam to the $|\pm1\rangle$ state to be near-resonant, i.e. $\Delta= \Delta_B-\omega_m \ll D_0$ as shown in Fig 1, the state $|0\rangle$ would remain unpopulated. In this way we can safely disregard it from our description. Under the rotating wave approximation each NV center can be viewed as a two-level system with an interaction Hamiltonian $H_i = g(\sigma_i^{+}a + a^\dagger\sigma_i^{-})$, where $\sigma_i^{\pm}=|\pm1\rangle_i\langle\mp|$ is the Pauli operator for the ith NV center. To describe our system where there are many NV centers in the nanoresonator, we introduce collective spin operators,$J_z = \frac{1}{2}\sum_i^N|1\rangle_i\langle1|-|-1\rangle_i\langle-1|$ and $J_{\pm}= J_x \pm iJ_y = \sum_i^N\sigma_i^{\pm}$ were $N$ is the total number of NV centers. It can be easily shown that those spin operators still obey the usual angular momentum commutation relations.  Thus the total Hamiltonian for the first ensemble indicated by superscript $1$ is
\begin{equation*}
H = \omega_m a^\dagger a + \Delta_B J_z^1 + g(a^\dagger J_{-}^1+aJ_{+}^1) \tag{2}
\end{equation*}

Using the Holstein-Primakoff transformation, we can map the collective spin operator into the bosonic operator c. If we prepare all the NVs in the first ensemble to be in $|-1\rangle$ state, the transformation is written as:
\begin{align*}
&J_z^1 = -\frac{1}{2}N + c_1^\dagger c_1\tag{3a}\\
&J_{-}^1 = c_1\sqrt{N-c_1^\dagger c_1}\simeq\sqrt{N}c_1 \tag{3b}\\
&J_{+}^1 = c_1^\dagger \sqrt{N_1c^\dagger c_1}\simeq\sqrt{N}c_1^\dagger\tag{3c}
\end{align*}

As long as the number of excited spins remains few, the ladder of state is well described by this approximation. And now the previous Hamiltonian (2) is:
\begin{equation*}
H_1 = \omega_m a^\dagger a + \Delta_{B1} (-\frac{1}{2}N+c_1^\dagger c_1) + \sqrt{N}g(a^\dagger c_1 + a c_1^\dagger) \tag{4}
\end{equation*}

While for the inverted NVE, everything goes all the same except that the NV centers are now in excited state. Because the Holstein-Primakoff approximation applies for small deviation from collectively occupied state\cite{kurucz2010multilevel}, the transformation here is slightly different from previous one:
\begin{align*}
&J_z^2 = \frac{1}{2}N - c_2^\dagger c_2\tag{5a}\\
&J_{-}^2 = c_2^\dagger\sqrt{N-c_2^\dagger c_2}\simeq\sqrt{N}c_2^\dagger\tag{5b}\\
&J_{+}^2 = c_2 \sqrt{N-_2c^\dagger c_2}\simeq\sqrt{N}c_2\tag{5c}
\end{align*}
Therefore the Hamiltonian for the inverted ensemble is
\begin{equation*}
H_2 = \omega_m b^\dagger b + \Delta_{B_2}( \frac{1}{2}N-c_2^\dagger c_2) + \sqrt{N}g(b^\dagger c_2^\dagger + b c_2) \tag{6}
\end{equation*}
where we use $b$ to represent the destruction operator for the mechanical mode in the second ensemble. We also assume same number of NVs in two ensembles and uniform coupling of each spin to the mechanical mode.

 The coupling strength and the frequency of mechanical mode can be calculated from Euler-Bernoulli thin beam elasticity theory if we take a doubly clamped diamond beam with $L \gg w,h$\cite{landau1986theory}. For NV centers near the surface of the beam, we have $\frac{g}{2\pi}\simeq 180\sqrt{\frac{\hbar}{L^3 w\sqrt{\rho E}}}$\cite{bennett2013phonon}. For a beam of dimensions (L,w,h) = (0.5,0.05,0.05)$\mu m $, we estimate the vibrational frequency $\omega_m/2\pi$$\sim$ 2 GHz and the coupling $g/2\pi$$\sim$4 kHz.

At last, we take the phonon-phonon interaction into consideration. When two NVEs are connected to each other as shown in Fig.1, the interaction Hamiltonian is written as  $H_3 = v(a^\dagger b + a b^\dagger)$ where v is the coupling strength of phonon-phonon interaction. In fact, this coupling can be adjusted by the control of how far this two ensembles are separated.

Now the total Hamiltonian of our scheme is $H = H_1+ H_2 +H_3$. We can see the mechanism of how this Hamiltonian creates the two-mode squeezed state. First the bosonic mode $c_1$ is entangled with the mechanical mode $a$ via $H_1$, then the two phonon are coupled together through the interaction process $H_3$, at last the entanglement of bosonic mode and the mechanical mode may be transfered to the bosonic mode $c_2$ via the linear mixing mechanism $H_2$. Therefore two-mode squeezed state of the separate NVE1 and NVE2 can be realized.
\section{Large detuning, adiabatic elimination of mechanical mode}

The bilinear Hamiltonian of our scheme leads to simple Heisenberg equations of motion for the oscillator ladder operator. In the following, we assume equal Zeeman splitting in two ensembles: $\Delta_{B1}$ = $\Delta_{B2}= \Delta$ and we redefine $g=10g$ to absorb the constant $\sqrt{N}$ where we set N=100. Then we go to the rotating frame at the frequency of $\omega_{nv}$, the equations of motion can be straightforwardly wrote down:
\begin{align*}
&\dot c_1 = - i g a\tag{7a}\\
&\dot c_2 = -i g b^\dagger\tag{7c}\\
&\dot a = i \omega a - i g c_1 - i v b \tag{7e}\\
&\dot b = i\omega b - i g c_2^\dagger - i v a \tag{7g}
\end{align*}
where $\omega= \Delta-\omega_{m}$ is the detuning, and the equations of motion for the Hermitian adjoint operators can be simply wrote down by taking the complex conjugation of the corresponding operators.

If the detuning $\omega$ which we set to be of $MHz$ is very large compared with g and the decay rate k which shall be shown to be smaller than g, then we can make the adiabatic elimination by setting the derivative of the mechanical mode to zero. We can understand this elimination better if we redefine two new operators $\tilde{a}$ and $\tilde{b}$ by $\tilde{a}= \frac{1}{\sqrt{2}}(a+b)$ and $\tilde{b}= \frac{1}{\sqrt{2}}(a-b)$, the coupling between two phonons will disappear:
\begin{align*}
&H = (w_{m}+v)\tilde{a}^\dagger \tilde{a} +(w_{m}-v)\tilde{b}^\dagger \tilde{b} + \Delta(c_1^\dagger c_1 - c_2^\dagger c_2)\\
&+ \frac{\sqrt{2}}{2}g[(\tilde{a}^\dagger +\tilde{b}^\dagger )c_1 + (\tilde{a}+\tilde{b})c_1^\dagger + (\tilde{a}^\dagger - \tilde{b}^\dagger)c_2^\dagger +(\tilde{a}-\tilde{b})c_2]\tag{8}
\end{align*}
Then the equations of motion for $\tilde{a}$ and $\tilde{b}$ are:
\begin{equation*}
\dot{\tilde{a}} = -i(\omega_{m}+v)\tilde{a} - i\frac{\sqrt{2}}{2}gc_1 - i\frac{\sqrt{2}}{2}gc_2^\dagger \tag{9}
\end{equation*}
\begin{equation*}
\dot{\tilde{b}} = -i(\omega_{m}-v)\tilde{b} - i\frac{\sqrt{2}}{2}gc_1 + i\frac{\sqrt{2}}{2}gc_2^\dagger \tag{10}
\end{equation*}

Notice that $\omega_m\pm v -\Delta = \omega \pm v $. So as long as this quantity is much larger than $g$ and $k$, we can adiabaticly eliminate $\tilde a$ and $\tilde b$. Mathematically speaking, this is just the same as to take the time derivative of a and b to be zero:
\begin{equation*}
0 = \dot a = i \omega a - i g c_1 - i v b \tag{11}
\end{equation*}
\begin{equation*}
0 = \dot b = i\omega b - i g c_2^\dagger - i v a \tag{12}
\end{equation*}
Solving these equations  we  find
\begin{equation*}
a = \frac{\omega g}{\omega^2-v^2}c_1 + \frac{v g}{\omega^2-v^2}c_2^\dagger \tag{13}
\end{equation*}
\begin{equation*}
b = \frac{\omega g}{\omega^2-v^2}c_2^\dagger +\frac{v g}{\omega^2-v^2}c_1 \tag{14}
\end{equation*}
And as for $a^\dagger$ or $b^\dagger$ we can just take the complex conjugation of $a$ or $b$.

To express the result more clearly, we define $A = \frac{\omega g^2}{w^2-v^2}$ and $B= \frac{vg^2}{w^2-v^2}$ and substitute the above results back to the equations of motion for $c_1$ and $c_2$, we can find the effective Hamiltonian:
\begin{equation*}
H_{eff} = A (c_1^\dagger c_1 + c_2^\dagger c_2) + B(c_1 c_2 + c_1^\dagger c_2^\dagger) \tag{15}
\end{equation*}

Notice the $c_1 c_2 + c_1^\dagger c_2^\dagger$  in the Hamiltonian which is very similar to the standard non-degenerate parametric amplifier Hamiltonian. As a matter of fact, this is a strong sign of two-mode squeezing. Now we can directly write down the equations of motion for $c_1$ and $c_2$
\begin{align*}
&\dot c_1 = - i A c_1 - i B c_2^\dagger \tag{16a} \\
&\dot c_1^\dagger = i A c_1^\dagger + i B c_2 \tag{16b}\\
&\dot c_2 = -i A c_2 - i B c_1^\dagger \tag{16c}\\
&\dot c_2^\dagger = i A c_2^\dagger + i B c_1 \tag{16d}
\end{align*}
Those equations can be straightforwardly solved, e.g,
\begin{align*}
c_1(t) = &\frac{e^{-i t \lambda}}{2 \lambda}[(-A e^{2 i t \lambda}+ \lambda e^{2 i t \lambda} + \lambda + A)c_1(0)\\
&+ B(1 - e^{2 i t \lambda})c_2^\dagger(0)] \tag{17}
\end{align*}
Where I have defined $\lambda = \sqrt{A^2-B^2}$

With those solutions, we can calculate the variance of quadrature phase operator:
\begin{equation*}
X_{c_1c_2}(\theta) = \frac{1}{2}(\frac{e^{i\theta}}{\sqrt{2}}(c_1+c_2) + \frac{e^{- i \theta}}{\sqrt{2}}(c_1^\dagger + c_2^\dagger)) \tag{18}
\end{equation*}
In the following we will focus on $X_{c_1c_2}(0)$. Since we started from the ground state, the expectation value of all the operators remains zero all the time, and we are left with the square of the quadrature phase operator to compute. Substitute the solutions to equations (16) to the expression of expression (18), we can find:
\begin{equation*}
V(X_{c_1 c_2}(0)) = \frac{1}{4}[1 - \frac{2v}{w+v}sin^2(2 \lambda t)] \tag{19}
\end{equation*}

 We can see that the squeezing is determined by the ratio of $\omega$ and v, namely the dutuning between the frequency of mechanical mode and  Zeeman splitting and the interaction strength of two mechanical mode. As a matter of fact, both values can be adjusted in our regime by manipulating the magnetic field strength and change the separation distance of the two resonators.

   In Fig.2(a) we show the variance of $X$ with different detunings. It is clearly that the closer the detuning is to the interaction strength, the larger the squeezing is. But they can't be the same, because our adiabatic elimination would become invalid otherwise. We also plotted several results in Fig.3 in which we numerically solved the original Hamiltonian without adiabatic elimination. From the figure we can see that our adiabatic approximation is indeed rather effective.
\begin{figure}[htbp]

\subfloat[]{
  \begin{centering}
    \includegraphics[width=1.62in]{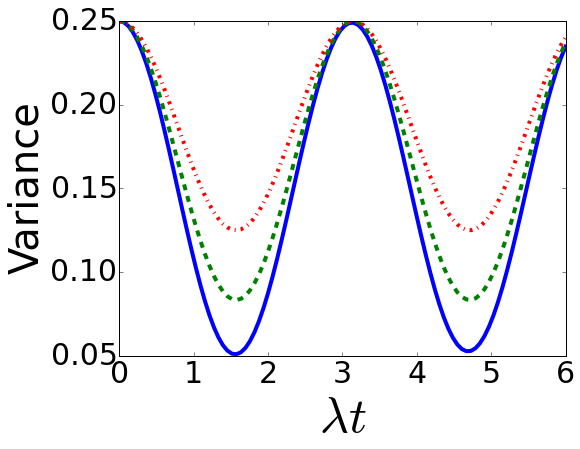}
\par\end{centering}
 }
 \subfloat[]{
    \begin{centering}
      \includegraphics[width=1.62in]{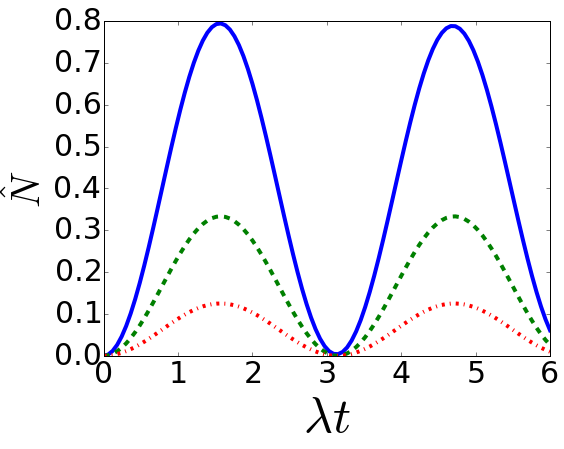}
      \par\end{centering}
}
\captionsetup{justification = raggedright}
\caption{\label{fig:2}
\begin{small}
(a) Squeezing of two NVEs under different detunings, we set g/$2\pi$ = 40KHz
and v/$2\pi$= 1MHz. Solid line shows the squeezing with $\omega= 1.5v$, dashed line represents
$\omega = 2v$, dashed doted line is for $\omega=3v$. (b)The number of excitations
under different situations. The style of line is the same as (a)
and we can see the number of excitations is much smaller than N=100 in our regime
which means our Hlstein-Primakoff approximation is very effective.
\end{small}  }
\end{figure}

\begin{figure}[htbp]
\begin{centering}
\subfloat[]{
  \begin{centering}
    \includegraphics[width=2.6in,height=1.6in]{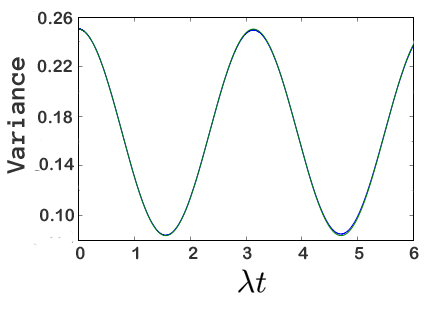}
\par\end{centering}
 }
 \vspace{0in}
\subfloat[]{
    \begin{centering}
      \includegraphics[width=2.6in,height=1.6in]{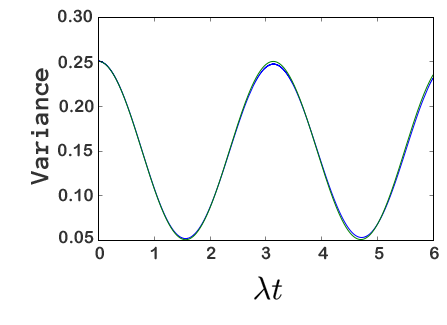}
      \par\end{centering}
}
\par\end{centering}
\captionsetup{justification = raggedright}
\label{fig:4}
\caption{We can see that the result obtained from the original
Hamiltonian is almost the same as the result we obtained using adiabatic elimination. The left figure shows the squeezing with $\omega = 2v$ and the right figure is for $\omega=1.5v$}
\end{figure}

At last we consider the Holstein-Primakoff approximation we made at the very beginning, the number of excited states is plotted in Fig.2(b) and we can see that, compared with the number of NV centers in our regime $N=100$, the excited number is indeed very few; thus the approximation is valid here.
\section{Different Zeeman splitting}
If the Zeeman splitting in two ensembles are different, the Hamiltonian of our system will take the form of :
\begin{align*}
H =&\omega_m(a^\dagger a + b^\dagger b)+\Delta_{B1}c_1^\dagger c_1 - \Delta_{B2}c_2^\dagger c_2\\
& + g(a^\dagger c_1+ a c_1^\dagger +b^\dagger c_2^\dagger +b c_2) + v(a^\dagger b+ab^\dagger) \tag{20}
\end{align*}
Go to the rotating system as before we can find the Heisenberg equations:
\begin{align*}
&\dot a = i\omega a -igc_1 -ivb \tag{21a}\\
&\dot b = i\omega b -igc_2^\dagger -ivb^\dagger \tag{21b}\\
&\dot c_1 = -iga \tag{21c}\\
&\dot c_2 = -i\Delta c_2 - igb^\dagger \tag{21d}
\end{align*}
where we have set $\omega = \Delta_{B1}-\omega_m$ and $\Delta = \Delta_{B1}-\Delta_{B2}$.
 If we assume further that $\Delta$ is not that big then we can still do the same adiabatic elimination and we find the effective Hamiltonian:
\begin{equation*}
H_eff= A c_1^\dagger c_1 +(A+\Delta)c_2^\dagger c_2 +B(c_1c_2 + c_1^\dagger c_2^\dagger) \tag{22}
\end{equation*}
where $A=\frac{\omega g^2}{\omega^2-v^2}$ and $B=\frac{v g^2}{\omega^2-v^2}$ as before. In this case, roughly speaking, when $\Delta$ falls outside of the range of $[-2A-2B,-2A+2B]$, the variance still shows oscillation feather. Indeed  we can improve our squeezing with the oscillation form solution. Several results are plotted in Figure. 4, from which we can see the decreased minimum squeezing compared with the result we obtained with same Zeeman splitting. But we can't tune it to reach the arbitrary minimum squeezing since this would bring larger number of excitations which would violate Hlstein-Primakoff approximation that we made at the very beginning.
\begin{figure}[htbp]

\subfloat[]{

    \includegraphics[width=1.6in,height=1.23in]{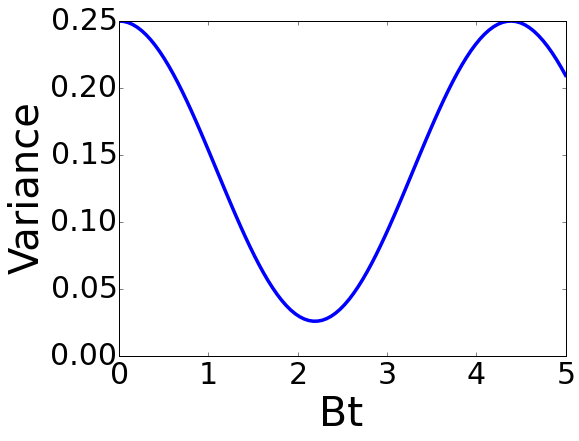}

 }
\subfloat[]{

      \includegraphics[width=1.6in,height=1.23in]{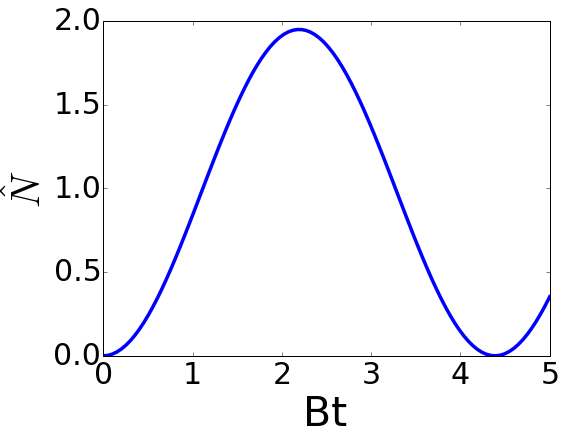}

}
\vspace{0in}
\subfloat[]{
\includegraphics[width = 1.6in,height=1.23in]{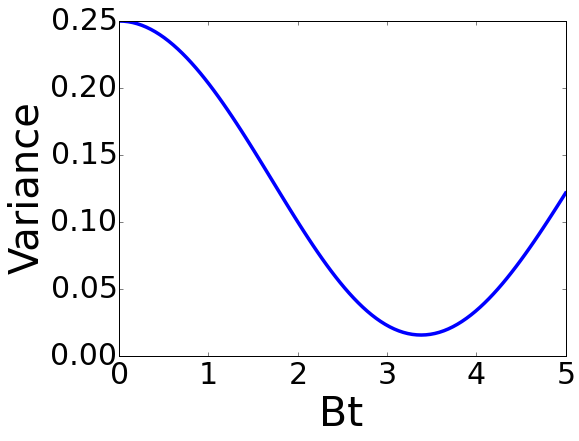}
    }
\subfloat[]{
\includegraphics[width = 1.6in,height=1.23in]{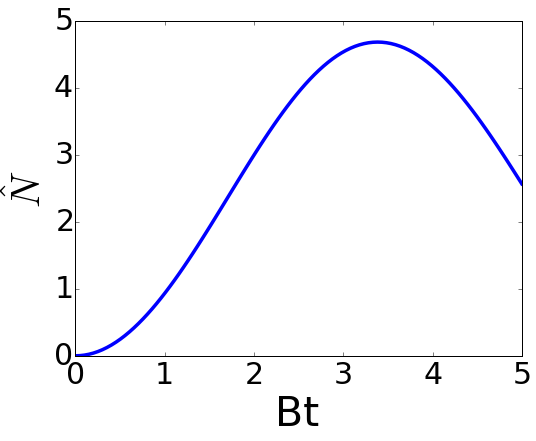}}
\captionsetup{justification = raggedright}
\label{fig:5}
\caption{(a) and (b) is the squeezing with $\Delta$= -0.77A while (c) and (d) is for $\Delta$= -0.9A. We set $g/2\pi = 40$ KHz and $v/2\pi =1$ MHz and $\omega = 2v$ As we can see (c) shows better squeezing but with larger excitations in our system which may cause our Holstein-Primakoff elimination less effective. On the other hand, (a) itself already shows much smaller squeezing compared with the previous result and the excitation number is still much small than 100.}
\end{figure}

If we investigate that effective Hamiltonian closer, we can find that in this situation,a special number operator $N_s= c_1^\dagger c_1 -c_2^\dagger c_2 $ remains a constant in Heisenberg picture due to the form of interaction Hamiltonian. Thus a special case would emerge from our system when $\Delta = -2A$, we can just take out the $A(c_1^\dagger c_1 - c_2^\dagger c_2)$ and we are left with:
\begin{equation*}
H_{eff} = B(c_1 c_2 + c_1^\dagger c_2^\dagger) \tag{23}
\end{equation*}
If we change the phase angle of our operator, i.e. redefine the operator: $c_1^\prime = e^{i\frac{\pi}{4}}c_1 $ and $c_2^\prime = e^{i\frac{\pi}{4}}c_2$. The effective Hamiltonian now is:
\begin{equation*}
H_{eff}= iB( c_1^{\prime\dagger} c_2^{\prime\dagger}- c_1^\prime c_2^\prime) \tag{24}
\end{equation*}
Actually this is just the standard unitary two-mode squeezing  operator in the interaction picture, the Heisenberg equations of motion are
\begin{align*}
&\dot c_1^\prime = Bc_2^{\prime\dagger}   ~~~  \dot c_2^{\prime\dagger} = B c_2^\prime \tag{25a}\\
&\dot c_2^\prime = B c_1^{\prime\dagger} ~~~ \dot c_1^{\prime\dagger} = B c_2^\prime \tag{25b}
\end{align*}
The solutions to these equations are
\begin{align*}
&c_1^\prime(t)=cosh(Bt)c_1^\prime(0) + sinh(Bt)c_2^{\prime\dagger}(0) \tag{26a}\\
&c_1^{\prime\dagger}(t) = cosh(Bt)c_1^{\prime\dagger}(0) + sinh(Bt)c_2^\prime(0) \tag{26b}\\
&c_2^\prime(t)= cosh(Bt) c_2^\prime(0) + sinh(Bt)c_1^{\prime\dagger} (0) \tag{26c}\\
&c_2^{\prime\dagger}(t) = cosh(Bt)c_2^{\prime\dagger}(0) + sinh(Bt)c_1^\prime(0) \tag{26d}
\end{align*}
Now we can calculate the variance of quadrature phase operator straightforwardly as before and the result is:
\begin{align*}
&V(X_{c_1^\prime c_2^\prime}(0)) =\frac{1}{4} e^{2Bt} \tag{27}\\
&V(X_{c_1^\prime c_2^\prime}(\frac{\pi}{2})) =\frac{1}{4} e^{-2Bt} \tag{27}
\end{align*}

If we let time goes to infinity, we expect perfectly correlated state which is the same as Einstein-Podolsky-Rosen state\cite{walls2007quantum}. But on the other hand, we can't expect this form of squeezing to last forever. As we can see from Fig.5, The number of excitations $c_1^\dagger c_1 = c_1^{\dagger\prime} c_1^{\prime} =sinh^2(Bt)$  can soon go beyond the limit where our Hlstein-Primakoff approximation becomes invalid.
\begin{figure}[htbp]
\begin{centering}

\subfloat[]{
    \begin{centering}
      \includegraphics[width=1.6in,height=1.25in]{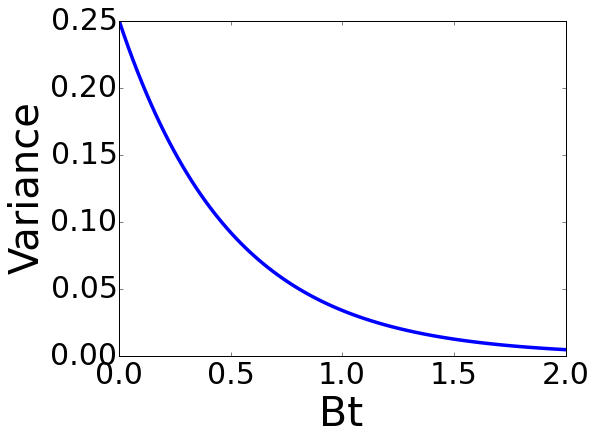}
      \par\end{centering}
}
\subfloat[]{
     \begin{centering}
       \includegraphics[width = 1.6in,height=1.25in]{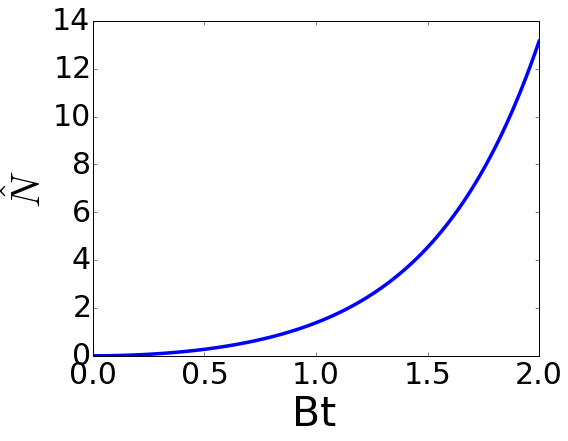}
       \par \end{centering}
}
\par\end{centering}
\captionsetup{justification = raggedright}
\label{fig:6}
\caption{(a)The variance of $ X_{c_1^\prime c_2^\prime}(\frac{\pi}{2})$, we set g/$2\pi$ = 40 KHz, v/$2\pi$= 1.5 MHz and $\omega= 2v$. (b) The number of excitations with same parameters. We can see our Halstein-Primakoff approximation will become less effective as time approach 1 ms.  }
\end{figure}
\section{Discussion}

Now we consider mechanical dissipation which is described by the master equation:
\begin{align*}
\dot \rho= &\kappa(\bar{n}_{th}+1)[a\rho a^\dagger -\frac{1}{2}(a^\dagger a \rho + \rho a^\dagger a)] + \kappa \bar{n}_{th}[a^\dagger \rho a \\
&- \frac{1}{2}(a a^\dagger\rho + \rho a a^\dagger)]+\kappa(\bar{n}_{th}+1)[b\rho b^\dagger -\frac{1}{2}(b^\dagger b \rho + \rho b^\dagger b)] \\
&+ \kappa \bar{n}_{th}[b^\dagger \rho b - \frac{1}{2}(b b^\dagger\rho + \rho b b^\dagger)], \tag{28}
\end{align*}
where $n_{th} = (e^{\hbar w_{ph}/\kappa_B T}-1)^{-1}$ is the equilibrium phonon occupation number at temperature T and $\kappa=\omega/Q$ is the mechanical damping rate. If we prepared our system at T=10 mK and assume $Q=10^6$ then we obtain $\kappa/2\pi = 1$ KHz\cite{tao2014single}. Because we started from the mechanical ground state and the adiabatic elimination remains effective in our regime, the average number of the phonon is far less than 1, and we can safely disregard the $k \bar{n}_{th}$ part in above description.

Roughly speaking, with the parameters that we set in the previous result, our system can reach to the minimum squeezing within half a millisecond. Besides, as we mentioned,the number of phonon is very few, we can safely draw the conclusion that the mechanical decay will not have any considerable effects on our result.

What's more, the electron spin relaxation time $T_1$ of NV centers in low temperature can reach several minutes\cite{jarmola2012temperature}. In addition, the dephasing time $T_2$ is also about several milliseconds at this low temperature\cite{balasubramanian2009ultralong}. Therefore these two rate should not seriously impact the result in our regime. .

In summary, we have proposed a scheme to generate two-mode squeezed state between two NVEs via the effective phonon-phonon interaction and phonon-spin interaction. The degree of squeezing can be manipulated; in a special case, our system can lead to a exponentially decreasing squeezing and we can reach considerable squeezing within the allowed time range. Besides, our structure is rather robust against all realistic conditions and is reachable with our currently available technology. From the future perspective, we can combine more nanoresonators together and form quantum network or quantum chains. With the phonons playing the role of quantum channels to transfer quantum information, we can expect the NVEs to be a good candidate to store, manipulate and process information.

This work is funded by the NBRPC (973 Program) 2011CBA00300 (2011CBA00301), NNSFC NO.61435007, NO. 11574353,
NO. 11274351 and NO.1105136.

\appendix
\section{Calculation of variance}
The equation (16) allows solutions which can by directly wrote down as:
\begin{align*}
c_1(t) = &\frac{e^{-i t \lambda}}{2 \lambda}[(-A e^{2 i t \lambda}+ \lambda e^{2 i t \lambda} + \lambda + A)c_1(0)\\
&+ B(1 - e^{2 i t \lambda})c_2^\dagger(0)] \tag{31a}
\end{align*}
\begin{align*}
c_2^\dagger(t) =  &\frac{e^{-i t \lambda}}{2 \lambda}[(A e^{2 i t \lambda}+ \lambda e^{2 i t \lambda} + \lambda - A)c_2^\dagger(0)\\
&+ B(e^{2 i t \lambda}-1)c_1(0)] \tag{31b}
\end{align*}
\begin{align*}
c_2(t) = &\frac{e^{-i t \lambda}}{2 \lambda}[(-A e^{2 i t \lambda}+ \lambda e^{2 i t \lambda} + \lambda + A)c_2(0)\\
&+ B(1 - e^{2 i t \lambda})c_1^\dagger(0)] \tag{31c}
\end{align*}
\begin{align*}
c_1^\dagger(t) = &\frac{e^{-i t \lambda}}{2 \lambda}[(A e^{2 i t \lambda}+ \lambda e^{2 i t \lambda} + \lambda - A)c_1^\dagger(0)\\
&+ B(e^{2 i t \lambda}-1)c_2(0)] \tag{31d}
\end{align*}
To compute the square of quadrature phase operator:
\begin{align*}
\langle X_{c_1 c_2}(0)^2 \rangle = &\frac{1}{8}\langle c_1c_1 + c_1c_2 + c_1c_1^\dagger + c_1c_2^\dagger + c_2c_1+ c_2c_2 \\
&+c_2c_1^\dagger + c_2c_2^\dagger+ c_1^\dagger c_1 + c_1^\dagger c_2 +c_1^\dagger c_1^\dagger\\
&+ c_1^\dagger c_2^\dagger+c_2^\dagger c_1 + c_2^\dagger c_2 + c_2^\dagger c_1^\dagger + c_2^\dagger c_2^\dagger \rangle \tag{32}
\end{align*}
We can set $\nu_1$ = $cos{t\lambda}- i\frac{A}{\lambda}sin£¨t\lambda£©$ and $\nu_2$ = $i\frac{B}{\lambda}sin(t\lambda)$, and remember that
$\langle c_1 c_1^\dagger(0) \rangle = \langle c_2 c_2^\dagger(0) \rangle = 1, \langle c_1^\dagger c_1(0) \rangle = \langle c_2^\dagger c_2(0) \rangle =0$, we can find these expressions using solutions in equation (31):
\begin{align*}
&\langle c_1 c_2\rangle = \langle c_2 c_1 \rangle = - \nu_1 \nu_2 \tag{33a} \\
&\langle c_1 c_1^\dagger \rangle = \langle c_2 c_2^\dagger \rangle = \nu_1 \nu_1^\dagger\tag{33b} \\
&\langle c_1^\dagger c_1 \rangle = \langle c_2^\dagger c_2 \rangle = - \nu_2^2 \tag{33c}\\
&\langle c_1^\dagger c_2^\dagger \rangle = \langle c_2^\dagger c_1^\dagger \rangle = \nu_2 \nu_1^\dagger \tag{33d}
\end{align*}
while other second moments are all zero.

Now the variance of X is just
\begin{equation*}
V(X_{c_1 c_2}(0)) = \frac{1}{4}( c_1 c_1^\dagger (t)+c_2^\dagger c_2 (t)+ c_1 c_2(t) + c_1^\dagger c_2^\dagger(t)) \tag{34}
\end{equation*}
 Put the previous results into the above expression, we can finally get the variance of X as a function of g w and v:
\begin{equation*}
V(X_{c_1 c_2}(0)) = \frac{1}{4}[1 - \frac{2v}{w+v}sin^2(2 \lambda t)] \tag{35}
\end{equation*}

\bibliography{citation}

\begin{thebibliography}{38}%
\makeatletter
\providecommand \@ifxundefined [1]{%
 \@ifx{#1\undefined}
}%
\providecommand \@ifnum [1]{%
 \ifnum #1\expandafter \@firstoftwo
 \else \expandafter \@secondoftwo
 \fi
}%
\providecommand \@ifx [1]{%
 \ifx #1\expandafter \@firstoftwo
 \else \expandafter \@secondoftwo
 \fi
}%
\providecommand \natexlab [1]{#1}%
\providecommand \enquote  [1]{``#1''}%
\providecommand \bibnamefont  [1]{#1}%
\providecommand \bibfnamefont [1]{#1}%
\providecommand \citenamefont [1]{#1}%
\providecommand \href@noop [0]{\@secondoftwo}%
\providecommand \href [0]{\begingroup \@sanitize@url \@href}%
\providecommand \@href[1]{\@@startlink{#1}\@@href}%
\providecommand \@@href[1]{\endgroup#1\@@endlink}%
\providecommand \@sanitize@url [0]{\catcode `\\12\catcode `\$12\catcode
  `\&12\catcode `\#12\catcode `\^12\catcode `\_12\catcode `\%12\relax}%
\providecommand \@@startlink[1]{}%
\providecommand \@@endlink[0]{}%
\providecommand \url  [0]{\begingroup\@sanitize@url \@url }%
\providecommand \@url [1]{\endgroup\@href {#1}{\urlprefix }}%
\providecommand \urlprefix  [0]{URL }%
\providecommand \Eprint [0]{\href }%
\providecommand \doibase [0]{http://dx.doi.org/}%
\providecommand \selectlanguage [0]{\@gobble}%
\providecommand \bibinfo  [0]{\@secondoftwo}%
\providecommand \bibfield  [0]{\@secondoftwo}%
\providecommand \translation [1]{[#1]}%
\providecommand \BibitemOpen [0]{}%
\providecommand \bibitemStop [0]{}%
\providecommand \bibitemNoStop [0]{.\EOS\space}%
\providecommand \EOS [0]{\spacefactor3000\relax}%
\providecommand \BibitemShut  [1]{\csname bibitem#1\endcsname}%
\let\auto@bib@innerbib\@empty
\bibitem [{\citenamefont {Bennett}\ \emph {et~al.}(2010)\citenamefont
  {Bennett}, \citenamefont {Cockins}, \citenamefont {Miyahara}, \citenamefont
  {Gr{\"u}tter},\ and\ \citenamefont {Clerk}}]{bennett2010strong}%
  \BibitemOpen
  \bibfield  {author} {\bibinfo {author} {\bibfnamefont {S.~D.}\ \bibnamefont
  {Bennett}}, \bibinfo {author} {\bibfnamefont {L.}~\bibnamefont {Cockins}},
  \bibinfo {author} {\bibfnamefont {Y.}~\bibnamefont {Miyahara}}, \bibinfo
  {author} {\bibfnamefont {P.}~\bibnamefont {Gr{\"u}tter}}, \ and\ \bibinfo
  {author} {\bibfnamefont {A.~A.}\ \bibnamefont {Clerk}},\ }\href@noop {}
  {\bibfield  {journal} {\bibinfo  {journal} {Phys. Rev. Lett.}\ }\textbf
  {\bibinfo {volume} {104}},\ \bibinfo {pages} {017203} (\bibinfo {year}
  {2010})}\BibitemShut {NoStop}%
\bibitem [{\citenamefont {Hammerer}\ \emph {et~al.}(2009)\citenamefont
  {Hammerer}, \citenamefont {Wallquist}, \citenamefont {Genes}, \citenamefont
  {Ludwig}, \citenamefont {Marquardt}, \citenamefont {Treutlein}, \citenamefont
  {Zoller}, \citenamefont {Ye},\ and\ \citenamefont
  {Kimble}}]{hammerer2009strong}%
  \BibitemOpen
  \bibfield  {author} {\bibinfo {author} {\bibfnamefont {K.}~\bibnamefont
  {Hammerer}}, \bibinfo {author} {\bibfnamefont {M.}~\bibnamefont {Wallquist}},
  \bibinfo {author} {\bibfnamefont {C.}~\bibnamefont {Genes}}, \bibinfo
  {author} {\bibfnamefont {M.}~\bibnamefont {Ludwig}}, \bibinfo {author}
  {\bibfnamefont {F.}~\bibnamefont {Marquardt}}, \bibinfo {author}
  {\bibfnamefont {P.}~\bibnamefont {Treutlein}}, \bibinfo {author}
  {\bibfnamefont {P.}~\bibnamefont {Zoller}}, \bibinfo {author} {\bibfnamefont
  {J.}~\bibnamefont {Ye}}, \ and\ \bibinfo {author} {\bibfnamefont {H.~J.}\
  \bibnamefont {Kimble}},\ }\href@noop {} {\bibfield  {journal} {\bibinfo
  {journal} {Phys. Rev. Lett.}\ }\textbf {\bibinfo {volume} {103}},\ \bibinfo
  {pages} {063005} (\bibinfo {year} {2009})}\BibitemShut {NoStop}%
\bibitem [{\citenamefont {Devoret}\ and\ \citenamefont
  {Schoelkopf}(2013)}]{devoret2013superconducting}%
  \BibitemOpen
  \bibfield  {author} {\bibinfo {author} {\bibfnamefont {M.}~\bibnamefont
  {Devoret}}\ and\ \bibinfo {author} {\bibfnamefont {R.}~\bibnamefont
  {Schoelkopf}},\ }\href@noop {} {\bibfield  {journal} {\bibinfo  {journal}
  {Science}\ }\textbf {\bibinfo {volume} {339}},\ \bibinfo {pages} {1169}
  (\bibinfo {year} {2013})}\BibitemShut {NoStop}%
\bibitem [{\citenamefont {Yin}\ \emph {et~al.}(2015)\citenamefont {Yin},
  \citenamefont {Zhao},\ and\ \citenamefont {Li}}]{yin2015hybrid}%
  \BibitemOpen
  \bibfield  {author} {\bibinfo {author} {\bibfnamefont {Z.}~\bibnamefont
  {Yin}}, \bibinfo {author} {\bibfnamefont {N.}~\bibnamefont {Zhao}}, \ and\
  \bibinfo {author} {\bibfnamefont {T.}~\bibnamefont {Li}},\ }\href@noop {}
  {\bibfield  {journal} {\bibinfo  {journal} {Sci China Phys Mech Astron}\
  }\textbf {\bibinfo {volume} {58}},\ \bibinfo {pages} {050303} (\bibinfo
  {year} {2015})}\BibitemShut {NoStop}%
\bibitem [{\citenamefont {{Rong}}\ \emph {et~al.}(2015)\citenamefont {{Rong}},
  \citenamefont {{Geng}}, \citenamefont {{Shi}}, \citenamefont {{Liu}},
  \citenamefont {{Xu}}, \citenamefont {{Ma}}, \citenamefont {{Kong}},
  \citenamefont {{Jiang}}, \citenamefont {{Wu}},\ and\ \citenamefont
  {{Du}}}]{Dujiangfeng}%
  \BibitemOpen
  \bibfield  {author} {\bibinfo {author} {\bibfnamefont {X.}~\bibnamefont
  {{Rong}}}, \bibinfo {author} {\bibfnamefont {J.}~\bibnamefont {{Geng}}},
  \bibinfo {author} {\bibfnamefont {F.}~\bibnamefont {{Shi}}}, \bibinfo
  {author} {\bibfnamefont {Y.}~\bibnamefont {{Liu}}}, \bibinfo {author}
  {\bibfnamefont {K.}~\bibnamefont {{Xu}}}, \bibinfo {author} {\bibfnamefont
  {W.}~\bibnamefont {{Ma}}}, \bibinfo {author} {\bibfnamefont {F.}~\bibnamefont
  {{Kong}}}, \bibinfo {author} {\bibfnamefont {Z.}~\bibnamefont {{Jiang}}},
  \bibinfo {author} {\bibfnamefont {Y.}~\bibnamefont {{Wu}}}, \ and\ \bibinfo
  {author} {\bibfnamefont {J.}~\bibnamefont {{Du}}},\ }\href@noop {} {\bibfield
   {journal} {\bibinfo  {journal} {Nat. Comm.}\ }\textbf {\bibinfo {volume}
  {6}},\ \bibinfo {pages} {8748} (\bibinfo {year} {2015})}\BibitemShut
  {NoStop}%
\bibitem [{\citenamefont {{Zu}}\ \emph {et~al.}(2014)\citenamefont {{Zu}},
  \citenamefont {{Wang}}, \citenamefont {{He}}, \citenamefont {{Zhang}},
  \citenamefont {{Dai}}, \citenamefont {{Wang}},\ and\ \citenamefont
  {{Duan}}}]{Zuchong}%
  \BibitemOpen
  \bibfield  {author} {\bibinfo {author} {\bibfnamefont {C.}~\bibnamefont
  {{Zu}}}, \bibinfo {author} {\bibfnamefont {W.-B.}\ \bibnamefont {{Wang}}},
  \bibinfo {author} {\bibfnamefont {L.}~\bibnamefont {{He}}}, \bibinfo {author}
  {\bibfnamefont {W.-G.}\ \bibnamefont {{Zhang}}}, \bibinfo {author}
  {\bibfnamefont {C.-Y.}\ \bibnamefont {{Dai}}}, \bibinfo {author}
  {\bibfnamefont {F.}~\bibnamefont {{Wang}}}, \ and\ \bibinfo {author}
  {\bibfnamefont {L.-M.}\ \bibnamefont {{Duan}}},\ }\href {\doibase
  10.1038/nature13729} {\bibfield  {journal} {\bibinfo  {journal} {\nat}\
  }\textbf {\bibinfo {volume} {514}},\ \bibinfo {pages} {72} (\bibinfo {year}
  {2014})}\BibitemShut {NoStop}%
\bibitem [{\citenamefont {Andersen}\ and\ \citenamefont
  {M{\o}lmer}(2012)}]{andersen2012squeezing}%
  \BibitemOpen
  \bibfield  {author} {\bibinfo {author} {\bibfnamefont {C.~K.}\ \bibnamefont
  {Andersen}}\ and\ \bibinfo {author} {\bibfnamefont {K.}~\bibnamefont
  {M{\o}lmer}},\ }\href@noop {} {\bibfield  {journal} {\bibinfo  {journal}
  {Phys. Rev A}\ }\textbf {\bibinfo {volume} {86}},\ \bibinfo {pages} {043831}
  (\bibinfo {year} {2012})}\BibitemShut {NoStop}%
\bibitem [{\citenamefont {Yang}\ \emph
  {et~al.}(2012{\natexlab{a}})\citenamefont {Yang}, \citenamefont {Yin},
  \citenamefont {Chen}, \citenamefont {Chen},\ and\ \citenamefont
  {Feng}}]{yang2012two}%
  \BibitemOpen
  \bibfield  {author} {\bibinfo {author} {\bibfnamefont {W.}~\bibnamefont
  {Yang}}, \bibinfo {author} {\bibfnamefont {Z.}~\bibnamefont {Yin}}, \bibinfo
  {author} {\bibfnamefont {Q.}~\bibnamefont {Chen}}, \bibinfo {author}
  {\bibfnamefont {C.}~\bibnamefont {Chen}}, \ and\ \bibinfo {author}
  {\bibfnamefont {M.}~\bibnamefont {Feng}},\ }\href@noop {} {\bibfield
  {journal} {\bibinfo  {journal} {Phys. Rev. A}\ }\textbf {\bibinfo {volume}
  {85}},\ \bibinfo {pages} {022324} (\bibinfo {year}
  {2012}{\natexlab{a}})}\BibitemShut {NoStop}%
\bibitem [{\citenamefont {Li}\ \emph {et~al.}(2012)\citenamefont {Li},
  \citenamefont {Gao}, \citenamefont {Li}, \citenamefont {Ma},\ and\
  \citenamefont {Li}}]{Lipengbo}%
  \BibitemOpen
  \bibfield  {author} {\bibinfo {author} {\bibfnamefont {P.-B.}\ \bibnamefont
  {Li}}, \bibinfo {author} {\bibfnamefont {S.-Y.}\ \bibnamefont {Gao}},
  \bibinfo {author} {\bibfnamefont {H.-R.}\ \bibnamefont {Li}}, \bibinfo
  {author} {\bibfnamefont {S.-L.}\ \bibnamefont {Ma}}, \ and\ \bibinfo {author}
  {\bibfnamefont {F.-L.}\ \bibnamefont {Li}},\ }\href {\doibase
  10.1103/PhysRevA.85.042306} {\bibfield  {journal} {\bibinfo  {journal} {Phys.
  Rev. A}\ }\textbf {\bibinfo {volume} {85}},\ \bibinfo {pages} {042306}
  (\bibinfo {year} {2012})}\BibitemShut {NoStop}%
\bibitem [{\citenamefont {Song}\ \emph {et~al.}(2015)\citenamefont {Song},
  \citenamefont {Yin}, \citenamefont {Yang}, \citenamefont {Zhu}, \citenamefont
  {Zhou},\ and\ \citenamefont {Feng}}]{song2015one}%
  \BibitemOpen
  \bibfield  {author} {\bibinfo {author} {\bibfnamefont {W.-l.}\ \bibnamefont
  {Song}}, \bibinfo {author} {\bibfnamefont {Z.-q.}\ \bibnamefont {Yin}},
  \bibinfo {author} {\bibfnamefont {W.-l.}\ \bibnamefont {Yang}}, \bibinfo
  {author} {\bibfnamefont {X.-b.}\ \bibnamefont {Zhu}}, \bibinfo {author}
  {\bibfnamefont {F.}~\bibnamefont {Zhou}}, \ and\ \bibinfo {author}
  {\bibfnamefont {M.}~\bibnamefont {Feng}},\ }\href@noop {} {\bibfield
  {journal} {\bibinfo  {journal} {Sci. Rep.}\ }\textbf {\bibinfo {volume} {5}}
  (\bibinfo {year} {2015})}\BibitemShut {NoStop}%
\bibitem [{\citenamefont {Yang}\ \emph {et~al.}(2011)\citenamefont {Yang},
  \citenamefont {Yin}, \citenamefont {Hu}, \citenamefont {Feng},\ and\
  \citenamefont {Du}}]{Yangwanli}%
  \BibitemOpen
  \bibfield  {author} {\bibinfo {author} {\bibfnamefont {W.~L.}\ \bibnamefont
  {Yang}}, \bibinfo {author} {\bibfnamefont {Z.~Q.}\ \bibnamefont {Yin}},
  \bibinfo {author} {\bibfnamefont {Y.}~\bibnamefont {Hu}}, \bibinfo {author}
  {\bibfnamefont {M.}~\bibnamefont {Feng}}, \ and\ \bibinfo {author}
  {\bibfnamefont {J.~F.}\ \bibnamefont {Du}},\ }\href {\doibase
  10.1103/PhysRevA.84.010301} {\bibfield  {journal} {\bibinfo  {journal} {Phys.
  Rev. A}\ }\textbf {\bibinfo {volume} {84}},\ \bibinfo {pages} {010301}
  (\bibinfo {year} {2011})}\BibitemShut {NoStop}%
\bibitem [{\citenamefont {Yang}\ \emph
  {et~al.}(2012{\natexlab{b}})\citenamefont {Yang}, \citenamefont {Yin},
  \citenamefont {Chen}, \citenamefont {Kou}, \citenamefont {Feng},\ and\
  \citenamefont {Oh}}]{Yangwanli1}%
  \BibitemOpen
  \bibfield  {author} {\bibinfo {author} {\bibfnamefont {W.~L.}\ \bibnamefont
  {Yang}}, \bibinfo {author} {\bibfnamefont {Z.-q.}\ \bibnamefont {Yin}},
  \bibinfo {author} {\bibfnamefont {Z.~X.}\ \bibnamefont {Chen}}, \bibinfo
  {author} {\bibfnamefont {S.-P.}\ \bibnamefont {Kou}}, \bibinfo {author}
  {\bibfnamefont {M.}~\bibnamefont {Feng}}, \ and\ \bibinfo {author}
  {\bibfnamefont {C.~H.}\ \bibnamefont {Oh}},\ }\href {\doibase
  10.1103/PhysRevA.86.012307} {\bibfield  {journal} {\bibinfo  {journal} {Phys.
  Rev. A}\ }\textbf {\bibinfo {volume} {86}},\ \bibinfo {pages} {012307}
  (\bibinfo {year} {2012}{\natexlab{b}})}\BibitemShut {NoStop}%
\bibitem [{\citenamefont {Ovartchaiyapong}\ \emph {et~al.}(2012)\citenamefont
  {Ovartchaiyapong}, \citenamefont {Pascal}, \citenamefont {Myers},
  \citenamefont {Lauria},\ and\ \citenamefont
  {Jayich}}]{ovartchaiyapong2012high}%
  \BibitemOpen
  \bibfield  {author} {\bibinfo {author} {\bibfnamefont {P.}~\bibnamefont
  {Ovartchaiyapong}}, \bibinfo {author} {\bibfnamefont {L.}~\bibnamefont
  {Pascal}}, \bibinfo {author} {\bibfnamefont {B.}~\bibnamefont {Myers}},
  \bibinfo {author} {\bibfnamefont {P.}~\bibnamefont {Lauria}}, \ and\ \bibinfo
  {author} {\bibfnamefont {A.~B.}\ \bibnamefont {Jayich}},\ }\href@noop {}
  {\bibfield  {journal} {\bibinfo  {journal} {Appl. Phys. Lett.}\ }\textbf
  {\bibinfo {volume} {101}},\ \bibinfo {pages} {163505} (\bibinfo {year}
  {2012})}\BibitemShut {NoStop}%
\bibitem [{\citenamefont {Zalalutdinov}\ \emph {et~al.}(2011)\citenamefont
  {Zalalutdinov}, \citenamefont {Ray}, \citenamefont {Photiadis}, \citenamefont
  {Robinson}, \citenamefont {Baldwin}, \citenamefont {Butler}, \citenamefont
  {Feygelson}, \citenamefont {Pate},\ and\ \citenamefont
  {Houston}}]{zalalutdinov2011ultrathin}%
  \BibitemOpen
  \bibfield  {author} {\bibinfo {author} {\bibfnamefont {M.~K.}\ \bibnamefont
  {Zalalutdinov}}, \bibinfo {author} {\bibfnamefont {M.~P.}\ \bibnamefont
  {Ray}}, \bibinfo {author} {\bibfnamefont {D.~M.}\ \bibnamefont {Photiadis}},
  \bibinfo {author} {\bibfnamefont {J.~T.}\ \bibnamefont {Robinson}}, \bibinfo
  {author} {\bibfnamefont {J.~W.}\ \bibnamefont {Baldwin}}, \bibinfo {author}
  {\bibfnamefont {J.~E.}\ \bibnamefont {Butler}}, \bibinfo {author}
  {\bibfnamefont {T.~I.}\ \bibnamefont {Feygelson}}, \bibinfo {author}
  {\bibfnamefont {B.~B.}\ \bibnamefont {Pate}}, \ and\ \bibinfo {author}
  {\bibfnamefont {B.~H.}\ \bibnamefont {Houston}},\ }\href@noop {} {\bibfield
  {journal} {\bibinfo  {journal} {Nano Lett.}\ }\textbf {\bibinfo {volume}
  {11}},\ \bibinfo {pages} {4304} (\bibinfo {year} {2011})}\BibitemShut
  {NoStop}%
\bibitem [{\citenamefont {Burek}\ \emph {et~al.}(2012)\citenamefont {Burek},
  \citenamefont {de~Leon}, \citenamefont {Shields}, \citenamefont {Hausmann},
  \citenamefont {Chu}, \citenamefont {Quan}, \citenamefont {Zibrov},
  \citenamefont {Park}, \citenamefont {Lukin},\ and\ \citenamefont
  {Lončar}}]{burek2012free}%
  \BibitemOpen
  \bibfield  {author} {\bibinfo {author} {\bibfnamefont {M.~J.}\ \bibnamefont
  {Burek}}, \bibinfo {author} {\bibfnamefont {N.~P.}\ \bibnamefont {de~Leon}},
  \bibinfo {author} {\bibfnamefont {B.~J.}\ \bibnamefont {Shields}}, \bibinfo
  {author} {\bibfnamefont {B.~J.}\ \bibnamefont {Hausmann}}, \bibinfo {author}
  {\bibfnamefont {Y.}~\bibnamefont {Chu}}, \bibinfo {author} {\bibfnamefont
  {Q.}~\bibnamefont {Quan}}, \bibinfo {author} {\bibfnamefont {A.~S.}\
  \bibnamefont {Zibrov}}, \bibinfo {author} {\bibfnamefont {H.}~\bibnamefont
  {Park}}, \bibinfo {author} {\bibfnamefont {M.~D.}\ \bibnamefont {Lukin}}, \
  and\ \bibinfo {author} {\bibfnamefont {M.}~\bibnamefont {Lončar}},\
  }\href@noop {} {\bibfield  {journal} {\bibinfo  {journal} {Nano Lett.}\
  }\textbf {\bibinfo {volume} {12}},\ \bibinfo {pages} {6084} (\bibinfo {year}
  {2012})}\BibitemShut {NoStop}%
\bibitem [{\citenamefont {Arcizet}\ \emph {et~al.}(2011)\citenamefont
  {Arcizet}, \citenamefont {Jacques}, \citenamefont {Siria}, \citenamefont
  {Poncharal}, \citenamefont {Vincent},\ and\ \citenamefont {Seidelin}}]{2011}%
  \BibitemOpen
  \bibfield  {author} {\bibinfo {author} {\bibfnamefont {O.}~\bibnamefont
  {Arcizet}}, \bibinfo {author} {\bibfnamefont {V.}~\bibnamefont {Jacques}},
  \bibinfo {author} {\bibfnamefont {A.}~\bibnamefont {Siria}}, \bibinfo
  {author} {\bibfnamefont {P.}~\bibnamefont {Poncharal}}, \bibinfo {author}
  {\bibfnamefont {P.}~\bibnamefont {Vincent}}, \ and\ \bibinfo {author}
  {\bibfnamefont {S.}~\bibnamefont {Seidelin}},\ }\href@noop {} {\bibfield
  {journal} {\bibinfo  {journal} {Nat. Phys.}\ }\textbf {\bibinfo {volume}
  {7}},\ \bibinfo {pages} {879} (\bibinfo {year} {2011})}\BibitemShut {NoStop}%
\bibitem [{\citenamefont {Barfuss}\ \emph {et~al.}(2015)\citenamefont
  {Barfuss}, \citenamefont {Teissier}, \citenamefont {Neu}, \citenamefont
  {Nunnenkamp},\ and\ \citenamefont {Maletinsky}}]{2015}%
  \BibitemOpen
  \bibfield  {author} {\bibinfo {author} {\bibfnamefont {A.}~\bibnamefont
  {Barfuss}}, \bibinfo {author} {\bibfnamefont {J.}~\bibnamefont {Teissier}},
  \bibinfo {author} {\bibfnamefont {E.}~\bibnamefont {Neu}}, \bibinfo {author}
  {\bibfnamefont {A.}~\bibnamefont {Nunnenkamp}}, \ and\ \bibinfo {author}
  {\bibfnamefont {P.}~\bibnamefont {Maletinsky}},\ }\href@noop {} {\bibfield
  {journal} {\bibinfo  {journal} {Nat. Phys.}\ }\textbf {\bibinfo {volume}
  {11}},\ \bibinfo {pages} {820} (\bibinfo {year} {2015})}\BibitemShut
  {NoStop}%
\bibitem [{\citenamefont {Yin}\ \emph {et~al.}(2013)\citenamefont {Yin},
  \citenamefont {Li}, \citenamefont {Zhang},\ and\ \citenamefont
  {Duan}}]{Yin2013}%
  \BibitemOpen
  \bibfield  {author} {\bibinfo {author} {\bibfnamefont {Z.-q.}\ \bibnamefont
  {Yin}}, \bibinfo {author} {\bibfnamefont {T.}~\bibnamefont {Li}}, \bibinfo
  {author} {\bibfnamefont {X.}~\bibnamefont {Zhang}}, \ and\ \bibinfo {author}
  {\bibfnamefont {L.~M.}\ \bibnamefont {Duan}},\ }\href {\doibase
  10.1103/PhysRevA.88.033614} {\bibfield  {journal} {\bibinfo  {journal} {Phys.
  Rev. A}\ }\textbf {\bibinfo {volume} {88}},\ \bibinfo {pages} {033614}
  (\bibinfo {year} {2013})}\BibitemShut {NoStop}%
\bibitem [{\citenamefont {Bennett}\ \emph {et~al.}(2013)\citenamefont
  {Bennett}, \citenamefont {Yao}, \citenamefont {Otterbach}, \citenamefont
  {Zoller}, \citenamefont {Rabl},\ and\ \citenamefont
  {Lukin}}]{bennett2013phonon}%
  \BibitemOpen
  \bibfield  {author} {\bibinfo {author} {\bibfnamefont {S.}~\bibnamefont
  {Bennett}}, \bibinfo {author} {\bibfnamefont {N.~Y.}\ \bibnamefont {Yao}},
  \bibinfo {author} {\bibfnamefont {J.}~\bibnamefont {Otterbach}}, \bibinfo
  {author} {\bibfnamefont {P.}~\bibnamefont {Zoller}}, \bibinfo {author}
  {\bibfnamefont {P.}~\bibnamefont {Rabl}}, \ and\ \bibinfo {author}
  {\bibfnamefont {M.~D.}\ \bibnamefont {Lukin}},\ }\href@noop {} {\bibfield
  {journal} {\bibinfo  {journal} {Phys. Rev. Lett.}\ }\textbf {\bibinfo
  {volume} {110}},\ \bibinfo {pages} {156402} (\bibinfo {year}
  {2013})}\BibitemShut {NoStop}%
\bibitem [{\citenamefont {MacQuarrie}\ \emph {et~al.}(2013)\citenamefont
  {MacQuarrie}, \citenamefont {Gosavi}, \citenamefont {Jungwirth},
  \citenamefont {Bhave},\ and\ \citenamefont
  {Fuchs}}]{macquarrie2013mechanical}%
  \BibitemOpen
  \bibfield  {author} {\bibinfo {author} {\bibfnamefont {E.}~\bibnamefont
  {MacQuarrie}}, \bibinfo {author} {\bibfnamefont {T.}~\bibnamefont {Gosavi}},
  \bibinfo {author} {\bibfnamefont {N.}~\bibnamefont {Jungwirth}}, \bibinfo
  {author} {\bibfnamefont {S.}~\bibnamefont {Bhave}}, \ and\ \bibinfo {author}
  {\bibfnamefont {G.}~\bibnamefont {Fuchs}},\ }\href@noop {} {\bibfield
  {journal} {\bibinfo  {journal} {Phys. Rev. Lett.}\ }\textbf {\bibinfo
  {volume} {111}},\ \bibinfo {pages} {227602} (\bibinfo {year}
  {2013})}\BibitemShut {NoStop}%
\bibitem [{\citenamefont {Albrecht}\ \emph {et~al.}(2013)\citenamefont
  {Albrecht}, \citenamefont {Retzker}, \citenamefont {Jelezko},\ and\
  \citenamefont {Plenio}}]{albrecht2013coupling}%
  \BibitemOpen
  \bibfield  {author} {\bibinfo {author} {\bibfnamefont {A.}~\bibnamefont
  {Albrecht}}, \bibinfo {author} {\bibfnamefont {A.}~\bibnamefont {Retzker}},
  \bibinfo {author} {\bibfnamefont {F.}~\bibnamefont {Jelezko}}, \ and\
  \bibinfo {author} {\bibfnamefont {M.~B.}\ \bibnamefont {Plenio}},\
  }\href@noop {} {\bibfield  {journal} {\bibinfo  {journal} {New J. Phys}\
  }\textbf {\bibinfo {volume} {15}},\ \bibinfo {pages} {083014} (\bibinfo
  {year} {2013})}\BibitemShut {NoStop}%
\bibitem [{\citenamefont {Goldman}\ \emph {et~al.}(2015)\citenamefont
  {Goldman}, \citenamefont {Sipahigil}, \citenamefont {Doherty}, \citenamefont
  {Yao}, \citenamefont {Bennett}, \citenamefont {Markham}, \citenamefont
  {Twitchen}, \citenamefont {Manson}, \citenamefont {Kubanek},\ and\
  \citenamefont {Lukin}}]{goldman2015phonon}%
  \BibitemOpen
  \bibfield  {author} {\bibinfo {author} {\bibfnamefont {M.}~\bibnamefont
  {Goldman}}, \bibinfo {author} {\bibfnamefont {A.}~\bibnamefont {Sipahigil}},
  \bibinfo {author} {\bibfnamefont {M.}~\bibnamefont {Doherty}}, \bibinfo
  {author} {\bibfnamefont {N.}~\bibnamefont {Yao}}, \bibinfo {author}
  {\bibfnamefont {S.}~\bibnamefont {Bennett}}, \bibinfo {author} {\bibfnamefont
  {M.}~\bibnamefont {Markham}}, \bibinfo {author} {\bibfnamefont
  {D.}~\bibnamefont {Twitchen}}, \bibinfo {author} {\bibfnamefont
  {N.}~\bibnamefont {Manson}}, \bibinfo {author} {\bibfnamefont
  {A.}~\bibnamefont {Kubanek}}, \ and\ \bibinfo {author} {\bibfnamefont
  {M.}~\bibnamefont {Lukin}},\ }\href@noop {} {\bibfield  {journal} {\bibinfo
  {journal} {Phys. Rev. Lett.}\ }\textbf {\bibinfo {volume} {114}},\ \bibinfo
  {pages} {145502} (\bibinfo {year} {2015})}\BibitemShut {NoStop}%
\bibitem [{\citenamefont {Macquarrie}\ \emph {et~al.}(2015)\citenamefont
  {Macquarrie}, \citenamefont {Gosavi}, \citenamefont {Moehle}, \citenamefont
  {Jungwirth}, \citenamefont {Bhave},\ and\ \citenamefont {Fuchs}}]{Optica}%
  \BibitemOpen
  \bibfield  {author} {\bibinfo {author} {\bibfnamefont {E.}~\bibnamefont
  {Macquarrie}}, \bibinfo {author} {\bibfnamefont {T.}~\bibnamefont {Gosavi}},
  \bibinfo {author} {\bibfnamefont {A.}~\bibnamefont {Moehle}}, \bibinfo
  {author} {\bibfnamefont {N.}~\bibnamefont {Jungwirth}}, \bibinfo {author}
  {\bibfnamefont {S.}~\bibnamefont {Bhave}}, \ and\ \bibinfo {author}
  {\bibfnamefont {G.}~\bibnamefont {Fuchs}},\ }\href@noop {} {\bibfield
  {journal} {\bibinfo  {journal} {Optica}\ }\textbf {\bibinfo {volume} {2}},\
  \bibinfo {pages} {233} (\bibinfo {year} {2015})}\BibitemShut {NoStop}%
\bibitem [{\citenamefont {Chen}\ \emph {et~al.}(2015)\citenamefont {Chen},
  \citenamefont {Liu}, \citenamefont {Peng}, \citenamefont {Zhi},\ and\
  \citenamefont {Xiao}}]{Xiaoyunfeng}%
  \BibitemOpen
  \bibfield  {author} {\bibinfo {author} {\bibfnamefont {X.}~\bibnamefont
  {Chen}}, \bibinfo {author} {\bibfnamefont {Y.-C.}\ \bibnamefont {Liu}},
  \bibinfo {author} {\bibfnamefont {P.}~\bibnamefont {Peng}}, \bibinfo {author}
  {\bibfnamefont {Y.}~\bibnamefont {Zhi}}, \ and\ \bibinfo {author}
  {\bibfnamefont {Y.-F.}\ \bibnamefont {Xiao}},\ }\href {\doibase
  10.1103/PhysRevA.92.033841} {\bibfield  {journal} {\bibinfo  {journal} {Phys.
  Rev. A}\ }\textbf {\bibinfo {volume} {92}},\ \bibinfo {pages} {033841}
  (\bibinfo {year} {2015})}\BibitemShut {NoStop}%
\bibitem [{\citenamefont {{Zhou}}\ \emph {et~al.}(2014)\citenamefont {{Zhou}},
  \citenamefont {{Hu}}, \citenamefont {{Yin}}, \citenamefont {{Wang}},
  \citenamefont {{Zhu}},\ and\ \citenamefont {{Xue}}}]{Xuezhengyuan}%
  \BibitemOpen
  \bibfield  {author} {\bibinfo {author} {\bibfnamefont {J.}~\bibnamefont
  {{Zhou}}}, \bibinfo {author} {\bibfnamefont {Y.}~\bibnamefont {{Hu}}},
  \bibinfo {author} {\bibfnamefont {Z.-Q.}\ \bibnamefont {{Yin}}}, \bibinfo
  {author} {\bibfnamefont {Z.~D.}\ \bibnamefont {{Wang}}}, \bibinfo {author}
  {\bibfnamefont {S.-L.}\ \bibnamefont {{Zhu}}}, \ and\ \bibinfo {author}
  {\bibfnamefont {Z.-Y.}\ \bibnamefont {{Xue}}},\ }\href {\doibase
  10.1038/srep06237} {\bibfield  {journal} {\bibinfo  {journal} {Scientific
  Reports}\ }\textbf {\bibinfo {volume} {4}},\ \bibinfo {eid} {6237} (\bibinfo
  {year} {2014})}\BibitemShut {NoStop}%
\bibitem [{\citenamefont {Li}\ \emph {et~al.}(2015)\citenamefont {Li},
  \citenamefont {Liu}, \citenamefont {Gao}, \citenamefont {Xiang},
  \citenamefont {Rabl}, \citenamefont {Xiao},\ and\ \citenamefont
  {Li}}]{Lipengbo1}%
  \BibitemOpen
  \bibfield  {author} {\bibinfo {author} {\bibfnamefont {P.-B.}\ \bibnamefont
  {Li}}, \bibinfo {author} {\bibfnamefont {Y.-C.}\ \bibnamefont {Liu}},
  \bibinfo {author} {\bibfnamefont {S.-Y.}\ \bibnamefont {Gao}}, \bibinfo
  {author} {\bibfnamefont {Z.-L.}\ \bibnamefont {Xiang}}, \bibinfo {author}
  {\bibfnamefont {P.}~\bibnamefont {Rabl}}, \bibinfo {author} {\bibfnamefont
  {Y.-F.}\ \bibnamefont {Xiao}}, \ and\ \bibinfo {author} {\bibfnamefont
  {F.-L.}\ \bibnamefont {Li}},\ }\href {\doibase
  10.1103/PhysRevApplied.4.044003} {\bibfield  {journal} {\bibinfo  {journal}
  {Phys. Rev. Applied}\ }\textbf {\bibinfo {volume} {4}},\ \bibinfo {pages}
  {044003} (\bibinfo {year} {2015})}\BibitemShut {NoStop}%
\bibitem [{\citenamefont {Maze}\ \emph {et~al.}(2011)\citenamefont {Maze},
  \citenamefont {Gali}, \citenamefont {Togan}, \citenamefont {Chu},
  \citenamefont {Trifonov}, \citenamefont {Kaxiras},\ and\ \citenamefont
  {Lukin}}]{maze2011properties}%
  \BibitemOpen
  \bibfield  {author} {\bibinfo {author} {\bibfnamefont {J.}~\bibnamefont
  {Maze}}, \bibinfo {author} {\bibfnamefont {A.}~\bibnamefont {Gali}}, \bibinfo
  {author} {\bibfnamefont {E.}~\bibnamefont {Togan}}, \bibinfo {author}
  {\bibfnamefont {Y.}~\bibnamefont {Chu}}, \bibinfo {author} {\bibfnamefont
  {A.}~\bibnamefont {Trifonov}}, \bibinfo {author} {\bibfnamefont
  {E.}~\bibnamefont {Kaxiras}}, \ and\ \bibinfo {author} {\bibfnamefont
  {M.}~\bibnamefont {Lukin}},\ }\href@noop {} {\bibfield  {journal} {\bibinfo
  {journal} {New J. Phys}\ }\textbf {\bibinfo {volume} {13}},\ \bibinfo {pages}
  {025025} (\bibinfo {year} {2011})}\BibitemShut {NoStop}%
\bibitem [{\citenamefont {Doherty}\ \emph {et~al.}(2012)\citenamefont
  {Doherty}, \citenamefont {Dolde}, \citenamefont {Fedder}, \citenamefont
  {Jelezko}, \citenamefont {Wrachtrup}, \citenamefont {Manson},\ and\
  \citenamefont {Hollenberg}}]{doherty2012theory}%
  \BibitemOpen
  \bibfield  {author} {\bibinfo {author} {\bibfnamefont {M.}~\bibnamefont
  {Doherty}}, \bibinfo {author} {\bibfnamefont {F.}~\bibnamefont {Dolde}},
  \bibinfo {author} {\bibfnamefont {H.}~\bibnamefont {Fedder}}, \bibinfo
  {author} {\bibfnamefont {F.}~\bibnamefont {Jelezko}}, \bibinfo {author}
  {\bibfnamefont {J.}~\bibnamefont {Wrachtrup}}, \bibinfo {author}
  {\bibfnamefont {N.}~\bibnamefont {Manson}}, \ and\ \bibinfo {author}
  {\bibfnamefont {L.}~\bibnamefont {Hollenberg}},\ }\href@noop {} {\bibfield
  {journal} {\bibinfo  {journal} {Phys. Rev. B}\ }\textbf {\bibinfo {volume}
  {85}},\ \bibinfo {pages} {205203} (\bibinfo {year} {2012})}\BibitemShut
  {NoStop}%
\bibitem [{\citenamefont {Holstein}\ and\ \citenamefont
  {Primakoff}(1940)}]{holstein1940field}%
  \BibitemOpen
  \bibfield  {author} {\bibinfo {author} {\bibfnamefont {T.}~\bibnamefont
  {Holstein}}\ and\ \bibinfo {author} {\bibfnamefont {H.}~\bibnamefont
  {Primakoff}},\ }\href@noop {} {\bibfield  {journal} {\bibinfo  {journal}
  {Phys. Rev.}\ }\textbf {\bibinfo {volume} {58}},\ \bibinfo {pages} {1098}
  (\bibinfo {year} {1940})}\BibitemShut {NoStop}%
\bibitem [{\citenamefont {Van~Oort}\ and\ \citenamefont
  {Glasbeek}(1990)}]{van1990electric}%
  \BibitemOpen
  \bibfield  {author} {\bibinfo {author} {\bibfnamefont {E.}~\bibnamefont
  {Van~Oort}}\ and\ \bibinfo {author} {\bibfnamefont {M.}~\bibnamefont
  {Glasbeek}},\ }\href@noop {} {\bibfield  {journal} {\bibinfo  {journal}
  {Chem. Phys. Lett.}\ }\textbf {\bibinfo {volume} {168}},\ \bibinfo {pages}
  {529} (\bibinfo {year} {1990})}\BibitemShut {NoStop}%
\bibitem [{\citenamefont {Dolde}\ \emph {et~al.}(2011)\citenamefont {Dolde},
  \citenamefont {Fedder}, \citenamefont {Doherty}, \citenamefont {N{\"o}bauer},
  \citenamefont {Rempp}, \citenamefont {Balasubramanian}, \citenamefont {Wolf},
  \citenamefont {Reinhard}, \citenamefont {Hollenberg}, \citenamefont {Jelezko}
  \emph {et~al.}}]{dolde2011electric}%
  \BibitemOpen
  \bibfield  {author} {\bibinfo {author} {\bibfnamefont {F.}~\bibnamefont
  {Dolde}}, \bibinfo {author} {\bibfnamefont {H.}~\bibnamefont {Fedder}},
  \bibinfo {author} {\bibfnamefont {M.~W.}\ \bibnamefont {Doherty}}, \bibinfo
  {author} {\bibfnamefont {T.}~\bibnamefont {N{\"o}bauer}}, \bibinfo {author}
  {\bibfnamefont {F.}~\bibnamefont {Rempp}}, \bibinfo {author} {\bibfnamefont
  {G.}~\bibnamefont {Balasubramanian}}, \bibinfo {author} {\bibfnamefont
  {T.}~\bibnamefont {Wolf}}, \bibinfo {author} {\bibfnamefont {F.}~\bibnamefont
  {Reinhard}}, \bibinfo {author} {\bibfnamefont {L.}~\bibnamefont
  {Hollenberg}}, \bibinfo {author} {\bibfnamefont {F.}~\bibnamefont {Jelezko}},
   \emph {et~al.},\ }\href@noop {} {\bibfield  {journal} {\bibinfo  {journal}
  {Nat. Phys.}\ }\textbf {\bibinfo {volume} {7}},\ \bibinfo {pages} {459}
  (\bibinfo {year} {2011})}\BibitemShut {NoStop}%
\bibitem [{\citenamefont {Acosta}\ \emph {et~al.}(2010)\citenamefont {Acosta},
  \citenamefont {Bauch}, \citenamefont {Ledbetter}, \citenamefont {Waxman},
  \citenamefont {Bouchard},\ and\ \citenamefont
  {Budker}}]{acosta2010temperature}%
  \BibitemOpen
  \bibfield  {author} {\bibinfo {author} {\bibfnamefont {V.}~\bibnamefont
  {Acosta}}, \bibinfo {author} {\bibfnamefont {E.}~\bibnamefont {Bauch}},
  \bibinfo {author} {\bibfnamefont {M.}~\bibnamefont {Ledbetter}}, \bibinfo
  {author} {\bibfnamefont {A.}~\bibnamefont {Waxman}}, \bibinfo {author}
  {\bibfnamefont {L.-S.}\ \bibnamefont {Bouchard}}, \ and\ \bibinfo {author}
  {\bibfnamefont {D.}~\bibnamefont {Budker}},\ }\href@noop {} {\bibfield
  {journal} {\bibinfo  {journal} {Phys. Rev. Lett.}\ }\textbf {\bibinfo
  {volume} {104}},\ \bibinfo {pages} {070801} (\bibinfo {year}
  {2010})}\BibitemShut {NoStop}%
\bibitem [{\citenamefont {Kurucz}\ and\ \citenamefont
  {M{\o}lmer}(2010)}]{kurucz2010multilevel}%
  \BibitemOpen
  \bibfield  {author} {\bibinfo {author} {\bibfnamefont {Z.}~\bibnamefont
  {Kurucz}}\ and\ \bibinfo {author} {\bibfnamefont {K.}~\bibnamefont
  {M{\o}lmer}},\ }\href@noop {} {\bibfield  {journal} {\bibinfo  {journal}
  {Phys. Rev. A}\ }\textbf {\bibinfo {volume} {81}},\ \bibinfo {pages} {032314}
  (\bibinfo {year} {2010})}\BibitemShut {NoStop}%
\bibitem [{\citenamefont {Landau}\ and\ \citenamefont
  {Lifshitz}(1986)}]{landau1986theory}%
  \BibitemOpen
  \bibfield  {author} {\bibinfo {author} {\bibfnamefont {L.}~\bibnamefont
  {Landau}}\ and\ \bibinfo {author} {\bibfnamefont {E.}~\bibnamefont
  {Lifshitz}},\ }\href@noop {} {\enquote {\bibinfo {title} {Theory of
  elasticity, 3rd},}\ } (\bibinfo {year} {1986})\BibitemShut {NoStop}%
\bibitem [{\citenamefont {Walls}\ and\ \citenamefont
  {Milburn}(2007)}]{walls2007quantum}%
  \BibitemOpen
  \bibfield  {author} {\bibinfo {author} {\bibfnamefont {D.~F.}\ \bibnamefont
  {Walls}}\ and\ \bibinfo {author} {\bibfnamefont {G.~J.}\ \bibnamefont
  {Milburn}},\ }\href@noop {} {\emph {\bibinfo {title} {Quantum optics}}}\
  (\bibinfo  {publisher} {Springer Science \& Business Media},\ \bibinfo {year}
  {2007})\BibitemShut {NoStop}%
\bibitem [{\citenamefont {Tao}\ \emph {et~al.}(2014)\citenamefont {Tao},
  \citenamefont {Boss}, \citenamefont {Moores},\ and\ \citenamefont
  {Degen}}]{tao2014single}%
  \BibitemOpen
  \bibfield  {author} {\bibinfo {author} {\bibfnamefont {Y.}~\bibnamefont
  {Tao}}, \bibinfo {author} {\bibfnamefont {J.}~\bibnamefont {Boss}}, \bibinfo
  {author} {\bibfnamefont {B.}~\bibnamefont {Moores}}, \ and\ \bibinfo {author}
  {\bibfnamefont {C.}~\bibnamefont {Degen}},\ }\href@noop {} {\bibfield
  {journal} {\bibinfo  {journal} {Nat. Commun.}\ }\textbf {\bibinfo {volume}
  {5}} (\bibinfo {year} {2014})}\BibitemShut {NoStop}%
\bibitem [{\citenamefont {Jarmola}\ \emph {et~al.}(2012)\citenamefont
  {Jarmola}, \citenamefont {Acosta}, \citenamefont {Jensen}, \citenamefont
  {Chemerisov},\ and\ \citenamefont {Budker}}]{jarmola2012temperature}%
  \BibitemOpen
  \bibfield  {author} {\bibinfo {author} {\bibfnamefont {A.}~\bibnamefont
  {Jarmola}}, \bibinfo {author} {\bibfnamefont {V.}~\bibnamefont {Acosta}},
  \bibinfo {author} {\bibfnamefont {K.}~\bibnamefont {Jensen}}, \bibinfo
  {author} {\bibfnamefont {S.}~\bibnamefont {Chemerisov}}, \ and\ \bibinfo
  {author} {\bibfnamefont {D.}~\bibnamefont {Budker}},\ }\href@noop {}
  {\bibfield  {journal} {\bibinfo  {journal} {Phys. Rev. Lett.}\ }\textbf
  {\bibinfo {volume} {108}},\ \bibinfo {pages} {197601} (\bibinfo {year}
  {2012})}\BibitemShut {NoStop}%
\bibitem [{\citenamefont {Balasubramanian}\ \emph {et~al.}(2009)\citenamefont
  {Balasubramanian}, \citenamefont {Neumann}, \citenamefont {Twitchen},
  \citenamefont {Markham}, \citenamefont {Kolesov}, \citenamefont {Mizuochi},
  \citenamefont {Isoya}, \citenamefont {Achard}, \citenamefont {Beck},
  \citenamefont {Tissler} \emph {et~al.}}]{balasubramanian2009ultralong}%
  \BibitemOpen
  \bibfield  {author} {\bibinfo {author} {\bibfnamefont {G.}~\bibnamefont
  {Balasubramanian}}, \bibinfo {author} {\bibfnamefont {P.}~\bibnamefont
  {Neumann}}, \bibinfo {author} {\bibfnamefont {D.}~\bibnamefont {Twitchen}},
  \bibinfo {author} {\bibfnamefont {M.}~\bibnamefont {Markham}}, \bibinfo
  {author} {\bibfnamefont {R.}~\bibnamefont {Kolesov}}, \bibinfo {author}
  {\bibfnamefont {N.}~\bibnamefont {Mizuochi}}, \bibinfo {author}
  {\bibfnamefont {J.}~\bibnamefont {Isoya}}, \bibinfo {author} {\bibfnamefont
  {J.}~\bibnamefont {Achard}}, \bibinfo {author} {\bibfnamefont
  {J.}~\bibnamefont {Beck}}, \bibinfo {author} {\bibfnamefont {J.}~\bibnamefont
  {Tissler}},  \emph {et~al.},\ }\href@noop {} {\bibfield  {journal} {\bibinfo
  {journal} {Nat. Mater}\ }\textbf {\bibinfo {volume} {8}},\ \bibinfo {pages}
  {383} (\bibinfo {year} {2009})}\BibitemShut {NoStop}%
\end{thebibliography}%

\end{document}